\documentclass[english,twocolumn,a4paper,svgnames]{article}
\usepackage[T1]{fontenc}
\usepackage[utf8]{inputenc}
\usepackage[english]{babel}
\usepackage{amsmath,amssymb,amsfonts,amsthm}
\usepackage{mathtools}
\providecommand{\ket}[1]{\left|#1\right\rangle}
\providecommand{\bra}[1]{\left\langle#1\right|}
\DeclareMathOperator{\Tr}{Tr}
\usepackage{graphicx}
\usepackage{subcaption}
\usepackage{tikz}
\usetikzlibrary{positioning,shapes.geometric,arrows,calc,matrix}
\usepackage{algorithmicx}
\usepackage{algpseudocode}
\usepackage{multirow}
\usepackage{hyperref}
\usepackage[toc,page]{appendix}
\usepackage{authblk}

\title{Non-Perturbative Topological Gadgets for Many-Body Coupling}
\author[1]{David Headley}
\author[1]{Nicholas Chancellor}
\affil[1]{School of Computing, Newcastle University, 1 Science Square,
Newcastle upon Tyne, NE4 5TG, UK}
\date{}

\begin{document}

\maketitle

\begin{abstract}
Continuous-time quantum hardware implementations generally lack the native capability to implement high-order terms that would facilitate efficient compilation of quantum algorithms. This limitation has, in part, motivated the development of \textit{perturbative} gadgets---multi-qubit constructions used to effect a desired Hamiltonian using engineered low-energy subspaces of a larger system constructed using simpler, usually two-body, primitives. In this work, we demonstrate how a class of \textit{non-perturbative} gadgets can produce high-order multi-body interactions by taking advantage of the odd-even properties of topological defect subspaces. 
The simplest example uses domain-wall defects in an effective Ising spin chain with linear connectivity and three-body couplings, alongside three- or five-body driving terms depending on the intended use. We demonstrate a version of a gadget which can perform an encoded bit-flip operation on a minor embedding chain, an important task to mitigate the limitations of quasi-two-dimensional (also sometimes called quasi-planar) topology. Although this will be the main focus of the paper due to conceptual simplicity, there exist systems constructed with only two-body couplings where the boundaries determine whether there are an odd or even number of defects, namely ice-like systems which may yield more complex gadget-like constructions.
\end{abstract}

\section{Introduction}

 The goal of this work is to produce gadgets for the implementation of multi-qubit Hamiltonians that are \textit{non-perturbative}. That is, they do not require either the use of perturbation theory to produce an effective Hamiltonian in a low-energy subspace or paths through penalized states to function, and therefore should not suffer from the exponentially large energy scales that are characteristic of perturbative gadgets. The domain-wall gadget studied here is the simplest example of a broader class of constructions based on boundary-constrained topological-defect sectors, expanded upon in Section~\ref{sec:extensions}. To obtain gadgets with such properties, we do, however, pay a price in \textit{composability}, in that using multiple gadgets of this form on shared groups of qubits cannot be expected to trivially work. We show, however, that useful tasks can still be accomplished with these designs. An extended discussion of perturbative gadgets and the associated literature can be found in Appendix~\ref{sec:perturbative_gadgets}.

The construction of these gadgets is possible through the use of encoded data-qubits with corresponding configurations of an ancillary system. This ancillary system carries locally, the global information as to the parity of the data qubits and therefore allows one to mediate an effective many-qubit interaction between data qubits. Importantly, these encoded states have finite overlap under certain accessible and local physical operations. The gadget presented in this work shares themes with the LHZ scheme, in which logical qubits are distributed over overlapping chains such that the parity information of qubit pairs becomes local and logical attributes, non-local, via the use of tiled four-body constraints~\cite{lechner2015quantum}. Likewise, in this gadget, the de-localization of the dressed chain states facilitates local access to the parity information of the whole string. In work extending the LHZ scheme to realistic transmon-based architectures~\cite{leib2016transmon}, three-body terms are used to construct required high-order penalties in a similar arrangement to this work. Related parity-based embedding architectures have also been proposed that avoid explicit $3$- and $4$-body constraints altogether. For example, Palacios et al.\ give an exclusively $2$-local construction for arbitrary-connectivity Ising models based on triangle decompositions and pairwise consistency constraints, albeit at the cost of substantial encoding overhead and the need for strong penalty terms to enforce consistency~\cite{Palacios2025TwoLocalParity}.

While gate-based quantum computers may freely compile multi-qubit gates through decompositions into a universal single- and two-qubit gate set, the same is not generally true for continuous-time, or hybrid, digital-analog quantum computers. The utility of many-body terms in continuous-time quantum computing is varied, with applications in whole chain driving of minor-embedded qubits~\cite{choi08a}, alongside that of driving logical qubits in adiabatic quantum error correction~\cite{pudenz2014error} and providing catalysts for speed-ups on some classes of optimization problems~\cite{ghosh2024exponential}. In search of the native capabilities to perform such tasks, the area of \textit{gadgets} has emerged, in which high-order terms are engineered from subspaces of larger, but lower-order interacting, Hamiltonians.
Hardware graph connectivity remains a persistent limitation in many quantum annealing architectures. This is evident in flux-qubit devices~\cite{Boothby2021Zephyr}: despite impressive gains in local graph connectivity, the global hardware graph remains quasi-two-dimensional, in the sense that each qubit can be assigned a two-dimensional neighbourhood containing all qubits to which it couples. Similar limitations are also present in emerging platforms, for example neutral atoms~\cite{Nguyen2023RydbergEmbedding,kombe2025quantumWire}.

Restricted connectivity changes the spin-glass properties of the resulting systems, generally making their ground states easier to find. Even when a problem is represented through minor embedding or parity schemes, there remains a fundamental tension between enforcing the constraints needed to represent a higher-dimensional problem faithfully and enabling the spin-flip dynamics needed to explore it. Encoded operations of the kind proposed in this work could remove that trade-off, because an encoded spin flip can be performed without ever violating a minor-embedding or parity constraint.

The paper is organized as follows. In Section~\ref{sec:conceptual_description}, we give a high-level description of the general structure underlying non-perturbative topological gadgets. In Section~\ref{sec:chain_gadget}, we introduce the domain-wall chain construction and the associated ancilla-driving terms. In Section~\ref{sec:forming_gadget}, we show how these ingredients combine to form a gapped gadget with an effective many-body interaction, and in Section~\ref{sec:encoded_operators}, we derive physical operators that implement useful dressed-basis operations. Section~\ref{section:performance_metrics} introduces the performance metrics used in our numerical analysis, while Section~\ref{sec:performance} presents numerical results demonstrating the action of driving on data qubits with imperfect confinement and the performance of the gadget in minor-embedding applications. In Section~\ref{sec:extensions}, we discuss possible extensions to more complex defect systems, and Section~\ref{sec:conclusion} concludes.

\section{Conceptual Description}
\label{sec:conceptual_description}

Before going into how the idea we propose here can be specifically implemented, it is worth first presenting the high-level concept. For this discussion, we assume the existence of Hamiltonians with eigenstates that have certain properties. The concrete example of the chain gadget discussed in Section~\ref{sec:chain_gadget} satisfies the necessary criteria, and we also discuss the possibility of others in Section~\ref{sec:extensions}. For this example, we assume that the goal is to impart a phase in the computational basis. This is equivalent to flipping many spins when translated to the $\ket{+},\ket{-}$ basis (as we do later), but we find it more pedagogical to discuss the construction in terms of phase.

To start with, let us consider a set of qubits which will be used to encode the state information of the system (we can call these data qubits). We will couple them to additional ancilla qubits, which we assume will remain in a unique ground state (by performing operations slowly enough relative to the strength of the Hamiltonian and relying on adiabatic effects), and therefore will effectively not increase the size of the Hilbert space used in the computation. Mathematically, the structure is as follows: each of the coupling terms with the ancillae takes the form
\begin{equation}
\hat{\mathcal{H}}_{\mathrm{coup}}(i)=Z_i\,\hat{\mathcal{H}}^{\rm a}_\mathrm{coup}(i)
\end{equation}
where we have not yet placed any restriction on the form of $\hat{\mathcal{H}}^{\rm a}_\mathrm{coup}(i)$.
Similarly, we assume that there is a part of the Hamiltonian that acts only on the ancillas, $\hat{\mathcal{H}}^{\rm a}$, so the total Hamiltonian is
\begin{equation}
\hat{\mathcal{H}}^\mathrm{tot}=\sum_i\hat{\mathcal{H}}_{\mathrm{coup}}(i)+\hat{\mathcal{H}}^{\rm a}
\label{eq:H_tot_abstract}\end{equation}
This Hamiltonian will be block-diagonal with blocks corresponding to each computational basis state (i.e.~every qubit in $\ket{0}$ or $\ket{1}$) of the data qubits. Writing the corresponding bitstring as $\mathbf{z}$ with components $z_i\in\{0,1\}$, we can write the $Z_i$ expectation value for the computational basis state $\mathbf{z}$ as 
\begin{equation}
\left \langle Z_i\right \rangle=(-1)^{z_i}\in\{1,-1\}, 
\end{equation}
where $z_i$ is the classical bit value encoded on data qubit $i$.
The total Hamiltonian then becomes
\begin{equation}
\hat{\mathcal{H}}^\mathrm{block}(\mathbf{z})=\sum_i(-1)^{z_i}\hat{\mathcal{H}}^{\rm a}_\mathrm{coup}(i)+\hat{\mathcal{H}}^{\rm a}
\end{equation}
This Hamiltonian can then be diagonalized, with a ground-state energy $E^{(0)}(\mathbf{z})$ and a ground state $\ket{\psi_{\mathbf{z},0}}$. We assume this Hamiltonian has a unique ground state. 

We now make the following observation: if $E^{(0)}(\mathbf{z})$ only depends on the parity of the sum of the bits, in other words
\begin{equation}
    E^{(0)}(\mathbf{z})=\begin{cases} 
    E^{(0)}_{\mathrm{odd}} & \sum_i z_i \quad \mathrm{odd} \\
    E^{(0)}_{\mathrm{even}} & \sum_i z_i \quad \mathrm{even}
    \end{cases},
\end{equation}
then, up to an additive constant and an unimportant global phase, the low-energy effective Hamiltonian on the data qubits is
\begin{equation}
    \hat{\mathcal{H}}_\mathrm{eff}=\frac{1}{2}\left(E^{(0)}_{\mathrm{even}}-E^{(0)}_{\mathrm{odd}}\right)\prod_i Z_i.
\end{equation}
We will show in later sections that such a structure can be guaranteed with topological defects, and moreover that this structure is local in the sense that the range between the minimum and maximum $i$ value of any term in Equation~\ref{eq:H_tot_abstract} is constant, independent of the size of the gadget.

For the gadgets proposed here to be useful, they have to also be able to be composed with other Hamiltonian terms. One of the major contributions of this work is to show that, at least in some cases, this can be done non-perturbatively. The crucial condition here is that the ancillas remain in their ground state while the encoded value changes. Such a transition between the state-encoding bitstring $\mathbf{z}$ and bitstring $\mathbf{z}'$ will generally be possible as long as\footnote{Note that Hamiltonian eigenstates are only defined up to a global phase so the phase of this inner product can be arbitrarily chosen.}
\begin{equation}
    \left| \left\langle\psi_{\mathbf{z},0}\middle| \psi_{\mathbf{z}',0} \right \rangle\right| >0. \label{eq:non_perturb_state_condition}
\end{equation}
Mathematically, nothing enforces orthogonality between pairs of states, and we show in later sections that in many important cases we can engineer Hamiltonians such that the necessary terms can be implemented non-perturbatively. For example, in the case of an encoded flip of a minor-embedding chain, single-qubit phases would need to be present, corresponding to single-bit flips in the basis we are using here. 

If we wish to add more terms which act between two bitstrings $\mathbf{z}$ and $\mathbf{z}'$, but do not have a convenient way to ensure that the condition in Equation~\ref{eq:non_perturb_state_condition} is met, we can perform the operation instead using
\begin{equation}
    \hat{\mathcal{H}}^\mathrm{enc}=\hat{\mathcal{H}}^\mathrm{sys}\hat{\mathcal{H}}^{\rm a}
\end{equation}
such that 
\begin{equation}
    \left| \left\langle\psi_{\mathbf{z},0}\middle| \hat{\mathcal{H}}^{\rm a} \middle| \psi_{\mathbf{z}',0} \right \rangle\right| >0. \label{eq:non_perturb_operator_condition}
\end{equation}
where $\hat{\mathcal{H}}^\mathrm{sys}$ acts on the data qubits to connect the computational-basis states $\mathbf{z}$ and $\mathbf{z}'$. We use this construction in Section~\ref{sub:two_enc} to implement two-body minor-embedding enforcing terms.

We have now shown the necessary ingredients for topological gadgets to function non-perturbatively in a completely abstract way, the rest of the manuscript will be devoted to concrete implementations which can achieve the mathematical structure described here.

\section{Chain Gadgets \label{sec:chain_gadget}}

\begin{figure*}
    \centering
\begin{tikzpicture}[
    font=\sffamily, % Default font for the picture
    data/.style={circle, draw, fill=black, minimum size=5mm, inner sep=0pt, font=\sffamily\bfseries}, % Text inside data nodes is bold
    ancilla/.style={circle, draw, fill=red!70, minimum size=5mm, inner sep=0pt, text=black, font=\sffamily\bfseries}, % Text inside ancilla nodes is bold
    interaction/.style={regular polygon, regular polygon sides=4, draw, fill=blue!50, minimum size=5mm, inner sep=1pt, text=black, font=\sffamily\bfseries}, % Text inside interaction nodes is bold
    virtual/.style={circle, draw, minimum size=7mm, inner sep=1pt, font=\sffamily\bfseries}, % Text inside virtual nodes is bold
    node distance=1.2cm and 1cm, % vertical and horizontal (vertical squished from 1.5cm)
    label distance=-2pt % Adjust distance of label from node edge for 'above'
]

% Virtual Ancilla Left
\node[virtual, label=above:{-1}] (VA-1) {$+1$}; % Added label above

% Ancilla 0
\node[ancilla, right=of VA-1] (A0) {};
% Interaction 0
\node[interaction, above=of A0] (H0) {};
% Data 0
\node[data, label=above:{0}, above=of H0] (D0) {};

% Connect first interaction
\draw (D0) -- (H0);
\draw (H0) -- (VA-1);
\draw (H0) -- (A0);

% Ancilla 1
\node[ancilla, right=of A0] (A1) {};
% Interaction 1
\node[interaction, above=of A1] (H1) {};
% Data 1
\node[data, label=above:{1}, above=of H1] (D1) {};

% Connect second interaction
\draw (D1) -- (H1);
\draw (H1) -- (A0);
\draw (H1) -- (A1);

% Ancilla 2
\node[ancilla, right=of A1] (A2) {};
% Interaction 2
\node[interaction, above=of A2] (H2) {};
% Data 2
\node[data, label=above:{2}, above=of H2] (D2) {};

% Connect third interaction
\draw (D2) -- (H2);
\draw (H2) -- (A1);
\draw (H2) -- (A2);

% Ancilla 3
\node[ancilla, right=of A2] (A3) {};
% Interaction 3
\node[interaction, above=of A3] (H3) {};
% Data 3
\node[data, label=above:{3}, above=of H3] (D3) {};

% Connect fourth interaction
\draw (D3) -- (H3);
\draw (H3) -- (A2);
\draw (H3) -- (A3);

% Ancilla 4
\node[ancilla, right=of A3] (A4) {};
% Interaction 4
\node[interaction, above=of A4] (H4) {};
% Data 4
\node[data, label=above:{4}, above=of H4] (D4) {};

% Connect fifth interaction
\draw (D4) -- (H4);
\draw (H4) -- (A3);
\draw (H4) -- (A4);

% Dots for continuation
\node[right=0.5cm of D4] (dotsD) {$\dots$};
\node[right=0.5cm of H4] (dotsH) {}; % Invisible node for positioning
\node[right=0.5cm of A4] (dotsA) {$\dots$};
% \draw[loosely dotted] (D4) -- (dotsD);
\draw[loosely dotted] (A4) -- ++(.7cm, .7cm);

% Last few elements (n-2, n-1)
% Data n-2
\node[data, label=above:{n-2}, right=0.5cm of dotsD] (Dn_2) {};
% Interaction n-2
\node[interaction, below=of Dn_2] (Hn_2) {};
% Ancilla n-2
\node[ancilla, below=of Hn_2] (An_2) {};

% Connect n-2 interaction
\draw (Dn_2) -- (Hn_2);
\draw (Hn_2) -- (An_2);
\draw[loosely dotted] ($(Hn_2) + (-0.7cm, -0.7cm)$) -- (Hn_2);
% Data n-1
\node[data, label=above:{n-1}, right=of Dn_2] (Dn_1) {};
% Interaction n-1
\node[interaction, below=of Dn_1] (Hn_1) {};

\node[virtual, label=above:{}, below=of Hn_1] (An_1_becomes_virtual) {$\pm 1$}; % Added label above

% Connect n-1 interaction
\draw (Dn_1) -- (Hn_1);
\draw (Hn_1) -- (An_2); % Connects to ancilla n-2
\draw (Hn_1) -- (An_1_becomes_virtual);  % Connects to the (n-1)th ancilla, which is now virtual

\end{tikzpicture}\hspace*{0.5cm}\begin{tikzpicture}[
    black_circle/.style={circle, draw=black, fill=black, minimum size=6pt, inner sep=0pt},
    blue_square/.style={rectangle, draw=black, fill=blue!50, minimum size=6pt, inner sep=1pt},
    red_circle/.style={circle, draw=black, fill=red!60, minimum size=6pt, inner sep=0pt},
    white_circle/.style={circle, draw=black, fill=white, minimum size=6pt, inner sep=0pt},
    line_style/.style={draw=black}
]

\matrix [draw, matrix of nodes, row sep=5pt, column sep=5pt, nodes={anchor=west}]
{
    \node[black_circle] {}; & \node {Data qubit}; \\
    \node[blue_square] {}; & \node {Term}; \\
    \node[red_circle] {}; & \node {Ancilla qubit}; \\
    \node[white_circle] {}; & \node {Boundary}; \\
    \draw[line_style] (0,0) -- (0.5,0); & \node {$Z$ term}; \\
};

\end{tikzpicture}
 \caption{A factor-graph representation of Equation~\ref{gadget_equation_fixed} in which blue vertices represent terms in the Hamiltonian, red vertices represent ancilla qubits and black vertices represent data qubits.\label{fig:factor_repr}} 
\end{figure*}

\paragraph*{Setup and Notation.}
A gadget, as used in this paper, is a system of qubits which can be used to implement a Hamiltonian that is not natively available, but can be engineered to be applied in a low-energy subspace of a larger qubit and ancilla system. The constituent components, in this work are defined as follows:
\begin{itemize}
    \item \( n_{\mathrm{d}} \) is the number of \textit{data} qubits (indexed as \( i = 0,1,\dots,n_{\mathrm{d}}-1 \)).
    \item \( n_{\mathrm{a}} = n_{\mathrm{d}} - 1\) is the number of \textit{ancilla} qubits (indexed as \( i = 0,1,\dots,n_{\mathrm{a}}-1 \)).
    \item $ \mathbf{z} = (z_0, \dots, z_{{n_{\rm d}}-1})$ and $ \quad \mathbf{a} = (a_0,\dots, a_{{n_{\rm a}}-1}) $ with \( z_i \in \{0,1\} \) for \( i = 0, \ldots, n_{\mathrm{d}}-1 \) and \( a_i \in \{0,1\} \) for \( i = 0, \ldots, n_{\mathrm{a}}-1 \) are binary-valued vectors describing computational basis states for data and ancilla qubits respectively.
    \item $\hat U_{\rm enc}$ is a unitary transformation which block diagonalizes the ancilla indices and encodes data qubit states with ancillae in the state $\ket{\bf {0}}$ to corresponding \textit{dressed} states.
    \item $\hat P$ is a projector into the subspace in which the ancillae are in the state $\ket{\bf {0}}$ in their block diagonalized basis with respect to a data-qubit state.
\end{itemize}

\paragraph*{Chain Construction.}
In this section, we introduce chain gadgets that make use of an overlapping sequence of three-body terms with linear connectivity. Later, we introduce driving terms that allow the use of these chains as gadgets rather than simply as large penalty terms. The form of the gadget shares features with that of gadgets of another work~\cite{Cichy_2024} in that it is constructed from overlapping three-body terms, however, with significant differences in the intended mode of operation and means of producing functional effective interactions. We define a \textit{chain} Hamiltonian, with boundary ancillae $Z^{\rm a}_{-1}$ and $Z^{\rm a}_{n_{\mathrm{d}}-1}$ understood to be fixed virtual variables, by:
\begin{equation}
    \hat{\mathcal{H}}_{\mathrm{Chain}} = \sum_{i=0}^{n_\mathrm{d}-1} \frac12\left(1-Z_{i}Z^{\rm a}_{i-1}Z^{\rm a}_{i}\right),
\end{equation}
which links the data and ancilla qubits in a contiguously overlapping chain of three-body parity-dependent terms.

We consider two implementations of the chain which implement identical interactions up to a change of sign. The first, \textit{unkinked} version has virtual ancillae fixed in their $+1$ eigenstates. The second, \textit{kinked} variant has the ancilla $n_\mathrm{d}-1$ pinned in its $-1$ $Z$ eigenstate, with the $-1$th ancilla still pinned in the $+1$ eigenstate. Substituting these virtual values, the Hamiltonian now takes the form:
\begin{multline}\label{gadget_equation_fixed}
    2\hat{\mathcal{H}}_{\mathrm{Chain}\pm} = \sum_{i=1}^{n_{\rm d}-2} \left(1-Z_{i}Z^{\rm a}_{i-1}Z^{\rm a}_{i}\right) \\
    +\left(1 - Z_0Z^{\rm a}_{0}\right) +\left(1 \mp Z_{n_{\rm d}-1}Z^{\rm a}_{n_{\rm d}-2}\right)
\end{multline}
which can be visualized in the form of a factor graph in Figure~\ref{fig:factor_repr}.

We wish to study the properties of the chain in the subspace where the ancilla qubits take a low-energy configuration relative to that of the data qubits. For a given data-qubit computational basis state $\mathbf{z}$, all terms in the chain Hamiltonian are satisfiable by an allocation of the ancilla qubits $\mathbf{a}$ if the parity of the data qubits is even (or odd) for the kinked (unkinked) chain Hamiltonian. If the Hamiltonian terms are satisfiable, the allocation $\mathbf{a}_\mathbf{z}$ can be uniquely determined. Iteratively, each term imposes that its co-indexed ancilla qubit takes a value equal to $Z^{\rm a}_{i} = Z^{\rm a}_{i-1}Z_i$ which determines:
\begin{equation}\label{satisfied}
        Z^{\rm a}_{i} = \prod_{i'=0}^{i} Z_{i'}.
\end{equation}
That is, ancillae carry the cumulative parity of the data qubits for increasing chain index.

In the case that the Hamiltonian is not satisfiable by an assignment of the ancillae, the low-energy subspace becomes degenerate, as any one of $n_\mathrm{d}$ terms may be chosen not to be satisfied. This introduces a \textit{topological defect} at position $j\in\{0,\ldots,n_{\mathrm{d}}-1\}$. 
For a given defect position $j$, the state of the ancillae can again be determined recursively, albeit with a \textit{kink} applied at index $j$, so
\begin{equation}
    Z^{\rm a}_{i}(j) = Z^{\rm a}_{i-1}(j)\,Z_i {(-1)}^{\delta(i,j)},
\end{equation}
and subsequently,
\begin{equation}
    Z^{\rm a}_{i}(j) = \prod_{i'=0}^{i} Z_{i'} {(-1)}^{\delta(i',j)},
\end{equation}
where $\delta(a,b)$ is the usual Kronecker delta function. In the frustrated case, the presence of \textit{domain walls}~\cite{Chancellor2019Domain} becomes apparent relative to the satisfiable case in that the ancillae are identical to their allocation given in Equation~\ref{satisfied}---the allocation matches until the presence of an unsatisfied clause, then takes the inverted allocation thereafter. Equation~\ref{satisfied} takes the role of the \textit{ice rule} for the chain, with the boundary conditions provided by the ancilla qubits determining whether the chain may be satisfied.

Substituting the fixed ancilla values into the full Hamiltonian, one obtains:
\begin{equation}
   \hat{\mathcal{H}}_\text{d} = \frac12 \left( 1 \mp \prod_{j=0}^{n_\mathrm{d}-1} Z_j \right)
\end{equation}
which demonstrates that the gadget imparts an energy consistent with the $n_{\rm d}$-body data-qubit parity, subject to the ancilla qubits remaining in their data-qubit-dependent ground state.

Such an effective Hamiltonian is useful if one wishes to penalize the dynamics of a set of qubits such that their parity remains odd or even, as in \cite{leib2016transmon}, but it is of limited utility as a term to be applied in more general contexts. This is because the excited ancilla states lie above the data-qubit-dependent ground-state subspace by the same energy separation that distinguishes odd and even data states, so this Hamiltonian cannot simultaneously apply a $Z^{n_{\rm d}}$ term and confine the ancillae to low-lying states. Therefore, one must add additional complexity to change the spectrum of the degenerate states such that odd-even dynamics occur in a subspace that is gapped from excited ancilla states. This is achieved by driving within the degenerate subspaces, breaking the degeneracy and providing a negative energy contribution where the data-qubit state does not allow zero-defect ancilla configurations.

\subsection{Driving Ancillae}

Thus far, the analysis has entirely followed a chain Hamiltonian as a classical spin system. In this section, we introduce conditional and non-conditional driving terms and investigate the resulting Hamiltonians. This driving is necessary because it adds an energetic contribution to the one-defect subspace, which, alongside the domain-wall penalty term, allows one to engineer both a gap between the zero- and one-defect subspaces and a larger gap between this dressed-state manifold and invalid ancilla configurations.

\subsubsection{Single-qubit driving}
Single-$X$ drivers induce transitions in the computational basis between states that differ by unit Hamming weight and are given by:
\begin{equation}
\hat{\mathcal{H}}_X = -\beta\sum_{i=0}^{n_{\rm d}-2} X^{\rm a}_i.
\end{equation}
The minus sign ensures that the low-energy configurations are smoothly varying states. Single-$X$ drivers have the advantage of being \textit{low weight} and, therefore, easy to physically implement on a device. However, they do not condition on the presence of existing defects and can therefore drive transitions that create or annihilate pairs of defects, rather than only moving existing defects. Representative examples of this action are shown in Figure~\ref{fig:single_x_defects}. Such drivers can be used, but with significant additional complexity in the form of a perturbative expansion or consideration of defect numbers greater than $1$. For work considering this in a gate-based setting, we refer the reader to work by Cichy et al.~\cite{Cichy_2024}.

\begin{figure*}
    \centering
\begin{tikzpicture}[
    font=\sffamily,
    data/.style={circle, draw, fill=black, minimum size=5mm, inner sep=0pt,
        font=\sffamily\bfseries, text=white},
    ancilla/.style={circle, draw, fill=red!70, minimum size=5mm, inner sep=0pt,
        font=\sffamily\bfseries, text=white},
    interaction/.style={regular polygon, regular polygon sides=4, draw, fill=blue!50,
        minimum size=5mm, inner sep=1pt, font=\sffamily\bfseries},
    brk/.style={regular polygon, regular polygon sides=4, draw=red!60!black,
        fill=orange!70, minimum size=5mm, inner sep=1pt, font=\sffamily\bfseries,
        line width=0.7pt},
    virtual/.style={circle, draw, minimum size=7mm, inner sep=1pt, font=\sffamily\bfseries},
    flipmark/.style={circle, draw=green!50!black, line width=1.5pt,
        dash pattern=on 2.5pt off 2pt, minimum size=8mm, inner sep=0pt},
    node distance=1.2cm and 1cm,
    label distance=-2pt,
    >=stealth,
]

\def\col{1}      % horizontal column separation
\def\row{1.2}    % vertical row separation
\def\ax{4.7}     % x-offset for "after" chain
\def\pb{-4.0}    % y-offset for panel (b)
\def\pc{-8.0}    % y-offset for panel (c)

\node[font=\sffamily\bfseries] at (-0.9, \row) {(a)};

\node[virtual, label=above:{-1}] (a-vl) at (0, 0) {$+1$};
\node[ancilla] (a-a0) at (\col, 0) {$+1$};
\node[interaction] (a-h0) at (\col, \row) {};
\node[data, label=above:{0}] (a-d0) at (\col, 2*\row) {$+1$};
\node[ancilla] (a-a1) at (2*\col, 0) {$+1$};
\node[interaction] (a-h1) at (2*\col, \row) {};
\node[data, label=above:{1}] (a-d1) at (2*\col, 2*\row) {$+1$};
\node[virtual] (a-vr) at (3*\col, 0) {$+1$};
\node[interaction] (a-h2) at (3*\col, \row) {};
\node[data, label=above:{2}] (a-d2) at (3*\col, 2*\row) {$+1$};
\draw (a-d0)--(a-h0) (a-h0)--(a-vl) (a-h0)--(a-a0);
\draw (a-d1)--(a-h1) (a-h1)--(a-a0) (a-h1)--(a-a1);
\draw (a-d2)--(a-h2) (a-h2)--(a-a1) (a-h2)--(a-vr);
\node[flipmark] at (a-a0) {};

\draw[->, thick] (3*\col+0.55, \row) --
    node[above] {$X^{\rm a}_0$} ++(0.8, 0);

\node[virtual, label=above:{-1}] (a2-vl) at (\ax, 0) {$+1$};
\node[ancilla] (a2-a0) at (\ax+\col, 0) {$-1$};
\node[brk] (a2-h0) at (\ax+\col, \row) {};
\node[data, label=above:{0}] (a2-d0) at (\ax+\col, 2*\row) {$+1$};
\node[ancilla] (a2-a1) at (\ax+2*\col, 0) {$+1$};
\node[brk] (a2-h1) at (\ax+2*\col, \row) {};
\node[data, label=above:{1}] (a2-d1) at (\ax+2*\col, 2*\row) {$+1$};
\node[virtual] (a2-vr) at (\ax+3*\col, 0) {$+1$};
\node[interaction] (a2-h2) at (\ax+3*\col, \row) {};
\node[data, label=above:{2}] (a2-d2) at (\ax+3*\col, 2*\row) {$+1$};
\draw (a2-d0)--(a2-h0) (a2-h0)--(a2-vl) (a2-h0)--(a2-a0);
\draw (a2-d1)--(a2-h1) (a2-h1)--(a2-a0) (a2-h1)--(a2-a1);
\draw (a2-d2)--(a2-h2) (a2-h2)--(a2-a1) (a2-h2)--(a2-vr);

\node[anchor=west, text width=3.2cm, align=left, font=\sffamily]
    at (\ax+3*\col+0.5, \row)
    {$0{\to}2$ defects:\\[1pt]
     $X^{\rm a}_0$ breaks two\\terms};

\node[font=\sffamily\bfseries] at (-0.9, \pb+\row) {(b)};

\node[virtual, label=above:{-1}] (b-vl) at (0, \pb) {$+1$};
\node[ancilla] (b-a0) at (\col, \pb) {$+1$};
\node[interaction] (b-h0) at (\col, \pb+\row) {};
\node[data, label=above:{0}] (b-d0) at (\col, \pb+2*\row) {$+1$};
\node[ancilla] (b-a1) at (2*\col, \pb) {$+1$};
\node[interaction] (b-h1) at (2*\col, \pb+\row) {};
\node[data, label=above:{1}] (b-d1) at (2*\col, \pb+2*\row) {$+1$};
\node[virtual] (b-vr) at (3*\col, \pb) {$+1$};
\node[brk] (b-h2) at (3*\col, \pb+\row) {};
\node[data, label=above:{2}] (b-d2) at (3*\col, \pb+2*\row) {$-1$};
\draw (b-d0)--(b-h0) (b-h0)--(b-vl) (b-h0)--(b-a0);
\draw (b-d1)--(b-h1) (b-h1)--(b-a0) (b-h1)--(b-a1);
\draw (b-d2)--(b-h2) (b-h2)--(b-a1) (b-h2)--(b-vr);
\node[flipmark] at (b-a0) {};

\draw[->, thick] (3*\col+0.55, \pb+\row) --
    node[above] {$X^{\rm a}_0$} ++(0.8, 0);

\node[virtual, label=above:{-1}] (b2-vl) at (\ax, \pb) {$+1$};
\node[ancilla] (b2-a0) at (\ax+\col, \pb) {$-1$};
\node[brk] (b2-h0) at (\ax+\col, \pb+\row) {};
\node[data, label=above:{0}] (b2-d0) at (\ax+\col, \pb+2*\row) {$+1$};
\node[ancilla] (b2-a1) at (\ax+2*\col, \pb) {$+1$};
\node[brk] (b2-h1) at (\ax+2*\col, \pb+\row) {};
\node[data, label=above:{1}] (b2-d1) at (\ax+2*\col, \pb+2*\row) {$+1$};
\node[virtual] (b2-vr) at (\ax+3*\col, \pb) {$+1$};
\node[brk] (b2-h2) at (\ax+3*\col, \pb+\row) {};
\node[data, label=above:{2}] (b2-d2) at (\ax+3*\col, \pb+2*\row) {$-1$};
\draw (b2-d0)--(b2-h0) (b2-h0)--(b2-vl) (b2-h0)--(b2-a0);
\draw (b2-d1)--(b2-h1) (b2-h1)--(b2-a0) (b2-h1)--(b2-a1);
\draw (b2-d2)--(b2-h2) (b2-h2)--(b2-a1) (b2-h2)--(b2-vr);

\node[anchor=west, text width=3.2cm, align=left, font=\sffamily]
    at (\ax+3*\col+0.5, \pb+\row)
    {$1{\to}3$ defects:\\[1pt]
     $X^{\rm a}_0$ breaks two\\additional terms};

\node[font=\sffamily\bfseries] at (-0.9, \pc+\row) {(c)};

\node[virtual, label=above:{-1}] (c-vl) at (0, \pc) {$+1$};
\node[ancilla] (c-a0) at (\col, \pc) {$+1$};
\node[interaction] (c-h0) at (\col, \pc+\row) {};
\node[data, label=above:{0}] (c-d0) at (\col, \pc+2*\row) {$+1$};
\node[ancilla] (c-a1) at (2*\col, \pc) {$+1$};
\node[interaction] (c-h1) at (2*\col, \pc+\row) {};
\node[data, label=above:{1}] (c-d1) at (2*\col, \pc+2*\row) {$+1$};
\node[virtual] (c-vr) at (3*\col, \pc) {$+1$};
\node[brk] (c-h2) at (3*\col, \pc+\row) {};
\node[data, label=above:{2}] (c-d2) at (3*\col, \pc+2*\row) {$-1$};
\draw (c-d0)--(c-h0) (c-h0)--(c-vl) (c-h0)--(c-a0);
\draw (c-d1)--(c-h1) (c-h1)--(c-a0) (c-h1)--(c-a1);
\draw (c-d2)--(c-h2) (c-h2)--(c-a1) (c-h2)--(c-vr);
\node[flipmark] at (c-a1) {};

\draw[->, thick] (3*\col+0.55, \pc+\row) --
    node[above] {$X^{\rm a}_1$} ++(0.8, 0);

\node[virtual, label=above:{-1}] (c2-vl) at (\ax, \pc) {$+1$};
\node[ancilla] (c2-a0) at (\ax+\col, \pc) {$+1$};
\node[interaction] (c2-h0) at (\ax+\col, \pc+\row) {};
\node[data, label=above:{0}] (c2-d0) at (\ax+\col, \pc+2*\row) {$+1$};
\node[ancilla] (c2-a1) at (\ax+2*\col, \pc) {$-1$};
\node[brk] (c2-h1) at (\ax+2*\col, \pc+\row) {};
\node[data, label=above:{1}] (c2-d1) at (\ax+2*\col, \pc+2*\row) {$+1$};
\node[virtual] (c2-vr) at (\ax+3*\col, \pc) {$+1$};
\node[interaction] (c2-h2) at (\ax+3*\col, \pc+\row) {};
\node[data, label=above:{2}] (c2-d2) at (\ax+3*\col, \pc+2*\row) {$-1$};
\draw (c2-d0)--(c2-h0) (c2-h0)--(c2-vl) (c2-h0)--(c2-a0);
\draw (c2-d1)--(c2-h1) (c2-h1)--(c2-a0) (c2-h1)--(c2-a1);
\draw (c2-d2)--(c2-h2) (c2-h2)--(c2-a1) (c2-h2)--(c2-vr);

\node[anchor=west, text width=3.2cm, align=left, font=\sffamily]
    at (\ax+3*\col+0.5, \pc+\row)
    {$1{\to}1$ defect:\\[1pt]
     $X^{\rm a}_1$ moves the\\broken term};

\end{tikzpicture}\hspace*{0.2cm}\raisebox{4.5cm}{%
\begin{tikzpicture}[
    black_circle/.style={circle, draw=black, fill=black, minimum size=6pt, inner sep=0pt},
    blue_square/.style={rectangle, draw=black, fill=blue!50, minimum size=6pt, inner sep=1pt},
    orange_square/.style={rectangle, draw=red!60!black, fill=orange!70, minimum size=6pt,
        inner sep=1pt, line width=0.5pt},
    red_circle/.style={circle, draw=black, fill=red!70, minimum size=6pt, inner sep=0pt},
    white_circle/.style={circle, draw=black, fill=white, minimum size=6pt, inner sep=0pt},
    flip_circle/.style={circle, draw=green!50!black, line width=1.5pt,
        dash pattern=on 1.5pt off 1pt,
        minimum size=10pt, inner sep=0pt},
    line_style/.style={draw=black},
]

\matrix [draw, matrix of nodes, row sep=5pt, column sep=5pt, nodes={anchor=west}, ampersand replacement=\&]
{
    \node[black_circle] {}; \& \node {Data qubit}; \\
    \node[blue_square] {};  \& \node {Satisfied term}; \\
    \node[red_circle] {};   \& \node {Ancilla qubit}; \\
    \node[white_circle] {}; \& \node {Boundary}; \\
    \draw[line_style] (0,0) -- (0.5,0); \& \node {$Z$ term}; \\
    \node[orange_square] {}; \& \node {Broken term}; \\
    \node[flip_circle] {};  \& \node {Qubit to flip}; \\
};

\end{tikzpicture}}
\caption{A factor-graph representation of the effect of applying single-qubit $X$ operations to ancilla qubits. (a) Applying $X^{\rm a}_i$ to a configuration in which all terms are satisfied breaks the two incident terms, creating a pair of defects. (b) If a single defect is already present and $X^{\rm a}_i$ acts on an ancilla qubit that is not a member of the broken term, two additional defects are created. (c) If a single defect is already present and $X^{\rm a}_i$ acts on an ancilla qubit in the broken term, the defect moves and the total number of defects is unchanged.}\label{fig:single_x_defects}
\end{figure*}

\subsubsection{Subspace drivers}

To drive the subspaces in which a single defect is present, we require a driver that applies an $X$ operator to an ancilla qubit, subject to a check that a defect is present at a connected site. As the ancilla is shared between adjacent terms, this will simultaneously flip the truth value of the two connected terms, moving the defect along the chain, or acting trivially in the no-defect case.

Therefore, we wish to construct a driver of the form:
\begin{multline}
\sum_{i=0}^{n_{\rm d}-2} \sum_{\mathbf{z}, \mathbf{a}} \ket{\textbf{z},\textbf{a}\oplus\mathbf{e}_i}\bra{\textbf{z}, \textbf{a}}  \\ \cdot
\left[\delta(i,j_\mathbf{z}(\mathbf{a})) + \delta(i+1,j_\mathbf{z}(\mathbf{a}))\right]
\end{multline}
where $j_\mathbf{z}(\mathbf{a})$ is the position of the broken clause implied by the allocations $\mathbf{z}$, $\mathbf{a}$, and $\mathbf{e}_i$ is the binary basis vector $(\mathbf{e}_i)_k = \delta_{ik}$, so that $\mathbf{a}\oplus\mathbf{e}_i$ flips ancilla $i$. We, therefore, need to find $Z$-based operators that have support on configurations where single defects are present. To do so, one must interrogate one clause or two adjacent clauses. To detect whether a defect occurs at position $j$, one can use the combination of operations $Z^{\rm a}_{i} Z_{i-1}^{\rm a} Z_i$ which carries the value $1$ if the defect is present and $-1$ otherwise.

Clauses can be chained with logical conjunctions. For example, the statement that a defect occurs at position $i$ \textsc{xor} $i+1$ is $Z_{i-1}^{\rm a} Z_{i}Z^{\rm a}_{i+1} Z_{i+1}$ where a cancellation has occurred at the $i$th position's ancilla qubit. To drive a transition between two defect states, we can drive another site conditional on there being a defect. We cannot drive an ancilla while simultaneously checking a non-commuting observable, and, therefore, checking a single broken clause cannot allow us to change the truth values of that clause. When checking two adjacent clauses, however, the shared ancilla qubit is independent of the \textsc{xor} of the two clauses. Therefore in this case the ancilla can be driven to move single defects.

As described, we can therefore define `clean' subspace mixers as:
\begin{equation}\label{equation_5bodydriver}
    \sum_{i=0}^{n_{\rm d} -2} X_i^{\rm a}\left[ 1 - Z_i Z_{i+1} Z^{\rm a}_{i-1} Z^{\rm a}_{i+1}\right].
\end{equation}
\begin{figure*}
    \centering
\begin{tikzpicture}[
    font=\sffamily,
    data/.style={circle, draw, fill=black, minimum size=5mm, inner sep=0pt, font=\sffamily\bfseries},
    ancilla/.style={circle, draw, fill=red!70, minimum size=5mm, inner sep=0pt, text=black, font=\sffamily\bfseries}, 
    interaction/.style={regular polygon, regular polygon sides=4, draw, fill=blue!50, minimum size=5mm, inner sep=1pt, text=black, font=\sffamily\bfseries},
    virtual/.style={circle, draw, minimum size=7mm, inner sep=1pt, font=\sffamily\bfseries},
    node distance=1.2cm and 1cm, 
    label distance=-2pt 
]

\node[virtual, label=above:{-1}] (VA-1) {$+1$};
% Ancilla 0
\node[ancilla, right=of VA-1] (A0) {};
% Interaction 0
\node[interaction, above=of A0] (H0) {};
% Data 0
\node[data, label=above:{0}, above=of H0] (D0) {};

% Connect first interaction
\draw[] (D0) -- (H0);
\draw[] (H0) -- (VA-1);
\draw[red] (H0) -- (A0);

% Ancilla 1
\node[ancilla, right=of A0] (A1) {};
% Interaction 1
\node[interaction, above=of A1] (H1) {};
% Data 1
\node[data, label=above:{1}, above=of H1] (D1) {};

\draw (A1) -- (H0);
\draw (D1) -- (H0);

% Connect second interaction
\draw (D1) -- (H1);
\draw (H1) -- (A0);
\draw[red] (H1) -- (A1);

% Ancilla 2
\node[ancilla, right=of A1] (A2) {};
% Interaction 2
\node[interaction, above=of A2] (H2) {};
% Data 2
\node[data, label=above:{2}, above=of H2] (D2) {};

\draw (D2) -- (H1);

% Connect third interaction
\draw (D2) -- (H2);
\draw (H2) -- (A1);
\draw[red] (H2) -- (A2);

% Ancilla 3
\node[ancilla, right=of A2] (A3) {};
% Interaction 3
\node[interaction, above=of A3] (H3) {};
% Data 3
\node[data, label=above:{3}, above=of H3] (D3) {};

\draw (D3) -- (H2);

% Connect fourth interaction
\draw (D3) -- (H3);
\draw (H3) -- (A2);
\draw[red] (H3) -- (A3);

% Ancilla 4
\node[ancilla, right=of A3] (A4) {};
% Interaction 4
\node[interaction, above=of A4] (H4) {};
% Data 4
\node[data, label=above:{4}, above=of H4] (D4) {};

\draw (D4) -- (H3);

% Connect fifth interaction
\draw (D4) -- (H4);
\draw (H4) -- (A3);
\draw[red] (H4) -- (A4);

% Dots for continuation
\node[right=0.5cm of D4] (dotsD) {$\dots$};
\node[right=0.5cm of H4] (dotsH) {}; % Invisible node for positioning
\node[right=0.5cm of A4] (dotsA) {$\dots$};
% \draw[loosely dotted] (D4) -- (dotsD);
\draw[loosely dotted] (A4) -- ++(.7cm, .7cm);
\draw[loosely dotted] (H4) -- ++(.7cm, -    .7cm);
\draw[loosely dotted] (H4) -- ++(+.7cm, +    .7cm);

% Last few elements (n-2, n-1)
% Data n-2
\node[data, label=above:{n-2}, right=0.5cm of dotsD] (Dn_2) {};
% Interaction n-2
\node[interaction, below=of Dn_2] (Hn_2) {};
% Ancilla n-2
\node[ancilla, below=of Hn_2] (An_2) {};

\draw (Dn_2) -- (Hn_2);
\draw[red] (Hn_2) -- (An_2);
\draw[loosely dotted] ($(Hn_2) + (-0.7cm, -0.7cm)$) -- (Hn_2);
\draw[loosely dotted] ($(Hn_2) + (-0.7cm, +0.7cm)$) -- (Hn_2);
\draw[loosely dotted] ($(An_2) + (-0.7cm, +0.7cm)$) -- (An_2);
\draw[loosely dotted] ($(Dn_2) + (-0.7cm, -0.7cm)$) -- (Dn_2);

% Data n-1
\node[data, label=above:{n-1}, right=of Dn_2] (Dn_1) {};

\node[virtual, label=above:{}, below=of An_2, right=of An_2] (An_1_becomes_virtual) {$\pm 1$};

% Connect n-1 interaction
\draw (Dn_1) -- (Hn_2);
\draw (Hn_2) -- (An_1_becomes_virtual);

\draw (A2) -- (H1);
\draw (A3) -- (H2);
\draw (A4) -- (H3);

\end{tikzpicture}\hspace*{0.2cm}
\begin{tikzpicture}[
    black_circle/.style={circle, draw=black, fill=black, minimum size=6pt, inner sep=0pt},
    blue_square/.style={rectangle, draw=black, fill=blue!50, minimum size=6pt, inner sep=1pt},
    red_circle/.style={circle, draw=black, fill=red!60, minimum size=6pt, inner sep=0pt},
    white_circle/.style={circle, draw=black, fill=white, minimum size=6pt, inner sep=0pt},
    line_style/.style={draw=black}
]

\matrix [draw, matrix of nodes, row sep=5pt, column sep=5pt, nodes={anchor=west}]
{
    \node[black_circle] {}; & \node {Data qubit}; \\
    \node[blue_square] {}; & \node {Term}; \\
    \node[red_circle] {}; & \node {Ancilla qubit}; \\
    \node[white_circle] {}; & \node {Boundary}; \\
    \draw[line_style] (0,0) -- (0.5,0); & \node {$Z$ term}; \\
    \draw[red] (0,0) -- (0.5,0); & \node {$X$ term}; \\
};
\end{tikzpicture}

    \caption{A factor-graph representation of Equation~\ref{equation_5bodydriver} in which blue vertices represent terms in the Hamiltonian, red vertices represent ancilla qubits and black vertices represent data qubits. The graph consists of overlapping five-body terms with driving on a single ancilla qubit dependent on $Z$ observables of locally accessible sites.\label{fig:factor_repr_subspace_driver}} 
\end{figure*}

The operator has the effect of linking one-defect states with values of $j = j_\mathbf{z}(\mathbf{a})$ differing by $1$. So, for a given value of $\mathbf{z}$ the effective applied Hamiltonian on the ancillae is:
\begin{equation}
   \hat{\mathcal{H}}_\mathrm{eff}= \sum_j 2 \Big[\ket{j}\bra{j+1} + \ket{j+1}\bra{j}\Big], 
\end{equation}
which is of tridiagonal Toeplitz form and is identical to the description of a particle in a discretized infinite square well, with the subspace driver taking the role of the kinetic energy term. Such operators are diagonalized by a sine transform. For a system with $n_{\rm d}$ basis states (representing the $n_{\rm d}$ possible defect positions, indexed $j = 0, \dots, n_{\rm d}-1$), the unitary matrix $\hat S$ effecting this transformation has elements $S_{kj}$ given by:
\begin{equation}\label{sine_transform_eqn}
    S_{kj} = \sqrt{\frac{2}{n_{\rm d}+1}} \sin\left(\frac{\pi (k+1)(j+1)}{n_{\rm d}+1}\right),
\end{equation}
where $k, j = 0, \dots, n_{\rm d}-1$. The columns of this matrix are the eigenvectors of $\hat{\mathcal{H}}_\mathrm{eff}$, with the eigenvalues taking the values:
\begin{equation}
\lambda_k = -2 \cos\left(\frac{\pi (k+1)}{n_{\rm d}+1}\right).
\end{equation}
The ground state of this effective Hamiltonian (corresponding to the eigenvalue $-2\cos(\frac{\pi}{n_{\rm d}+1})$) is the eigenvector 
\begin{equation}
    \sum_j S_{0j} \ket{j} =\sqrt{\frac{2}{n_{\rm d}+1}} \sum_j \sin\left(\frac{\pi (j+1)}{n_{\rm d}+1}\right) \ket{j}. 
\end{equation}
with the $k$th eigenvector taking the form \begin{equation}
\ket{\psi_k} = \sum_{j=0}^{n_{\rm d}-1} S_{kj} \ket{j}.
\end{equation}
Now, the chain-driver system can be diagonalized with a unitary $\hat U_\mathrm{enc}$ which diagonalizes the degenerate subspaces with a controlled application of $\hat S$ on the relevant data and ancilla subspaces. This takes the form:
\begin{equation}
\hat U_\text{enc} = \sum_{\mathbf{z},\mathbf{a}} \ket{\mathbf{z},\psi_{\mathbf{z},a}} \bra{\mathbf{z},\mathbf{a}},
\end{equation}
where $\psi_{\mathbf{z},a}$ is the $a$th excited state of the system projected into a computational basis state $\mathbf{z}$ of the data qubits, consisting of non-interacting subspaces with differing numbers of defects.

The subspace mixers described here are both five-local and of mixed types, the engineering of which on a physical device would be difficult even in comparison to the three-body terms forming the chain Hamiltonian. However, if no encoded operators are needed, as described in the following sections, one can make changes to the subspace drivers to reduce their order.

In Equation~\ref{equation_5bodydriver}, it is possible to multiply the expression by any operator without changing the property in which it has support only on the subspace with one incident defect. Therefore, one can multiply by any of the constituent $Z$ operators, changing the weight of the expression to $3$ or $4$. This, however, changes the form of the driver in the one-defect, or $j$ basis.

As such, one can construct another subspace mixer:
\begin{multline}
    \sum_{i=0}^{n_{\rm d}-2} X_i^{\rm a} \left[ Z_i Z_{i+1} - Z^{\rm a}_{i-1}Z_{i+1}^{\rm a} \right]
\\
=\sum_j 2Z_j Z_{j+1} \Big[\ket{j}\bra{j+1} + \ket{j+1}\bra{j}\Big].
\end{multline}
To transform this operator to Toeplitz form one must apply a gauge transform in which $\ket{j}\to g_j\ket{j}$, where $g_{j}g_{j+1} = Z_{j}Z_{j+1}$ which is trivially true for $g_j = Z_j$. The gauge unitary effecting this transformation is then:
\begin{equation}
    \hat U_\mathrm{Gauge} = \sum_\mathbf{z,a} {(-1)}^{z_{j_{\mathbf{z}(a)}}} \ket{\mathbf{z,a}}\bra{\mathbf{z,a}}
\end{equation}
which is an $n_{\rm d}$ qubit transformation. While commuting with the $Z$-based chain Hamiltonian, it will not commute with $X$-based operations on the data qubits, meaning that driving in the dressed basis will be complicated by the change of basis. This likely limits the applicability of the $3$-locally driven gadget outside the applications described in Section~\ref{sec:minor_embedding} concerning quantum adiabatic error correction or minor-chain embedding.

This section has presented the chain Hamiltonian and the driving Hamiltonians required to engineer the target many-body interaction. Crucially, while the ideal driving terms may be high order, the device requires at most five-local interactions regardless of gadget scale.

\section{Forming the Gadget}
\label{sec:forming_gadget}
The goal of this section is to combine the chain Hamiltonian and subspace drivers introduced in the previous section into a gapped gadget with a controllable effective many-body interaction.

\begin{figure*}
    \centering

    \includegraphics[width=\textwidth]{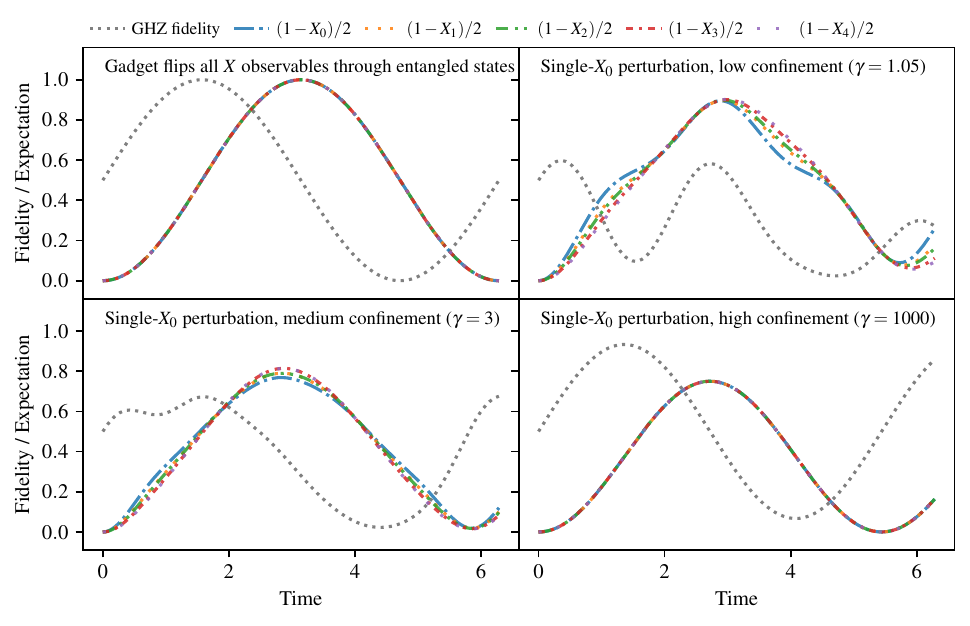}
    \vspace*{-.7cm}
    \caption{(a) The gadget Hamiltonian evolves a state initialized in the dressed $\ket{+}$ state such that all $X$ observables flip simultaneously and coherently. The time evolution produces a GHZ state for $t = \pi/2$, demonstrating that the effective Hamiltonian is the entangling $ZZ..Z$ operator. (b) The same system, with an added single-qubit physical $X$ term on the zeroth data qubit, causes different dynamics depending on the strength of the Hamiltonian confining the dynamics to the valid subspace. For low confinement (top right), the evolution of each dressed-basis $X$ observable becomes heterogeneous, indicating that the system is not adequately confined. For high values of confinement (bottom right), the system acts homogeneously on the dressed-basis $X$ observables, with dynamics that are isomorphic to those of a single-qubit Hamiltonian $\alpha X + Z/2$, in which $\alpha$ is the factor by which the dressed-basis $X$ operator is weaker than its physical counterpart due to non-unit overlap between the dressed states under the action of the physical operator.\label{fig:ghx_and_x_flips}}
\end{figure*}

So far, we have demonstrated a chain Hamiltonian $\gamma \hat{\mathcal{H}}_{\rm Chain}$ that penalizes the number of defects and subspace driver $\hat{\mathcal{H}}_{\rm Subspace}$ that adds a negative energy contribution of $2\cos(\frac{\pi}{n_{\rm d}+1})$ to the degenerate subspaces. The sum of these terms is:
\begin{equation}
    \hat{\mathcal{H}}_{ \rm Gadget} = \gamma \hat{\mathcal{H}}_{\rm Chain} + \beta\hat{\mathcal{H}}_{\rm Subspace}.
\end{equation}
We aim to produce a low-energy subspace in which the effective interaction $\alpha Z^{\otimes n_{\rm d}}$ is applied, with transitions out of this subspace discouraged with a large energy gap, therefore we should have $\gamma \gg\alpha$. To achieve a term strength $\alpha$ we can set:
\begin{equation}
    \beta = \frac{\gamma - \alpha}{ 2\cos\frac{\pi}{n_{\rm d} + 1}}
\end{equation}
which results in
\begin{equation}
    \hat{\mathcal{H}}_{\rm eff} = \hat P \hat U_\mathrm{enc}^\dagger  \hat{\mathcal{H}}_{ \rm Gadget}\hat U_\mathrm{enc}\hat P = \alpha Z^{\otimes n_{\rm d}}
\end{equation}
and a gap of size $O(\gamma)$ between this subspace and the next excited state. We informally refer to $\gamma$ as the confinement strength of the gadget, in that it determines the degree to which the dynamics are confined to this space. To demonstrate the action of the gadget asserted in this section, Figure~\ref{fig:ghx_and_x_flips} demonstrates in four panels the action of the gadget on qubit observables. In the first panel, we demonstrate that the gadget can start in an unentangled (dressed) $\ket{+}^{\otimes n_{\rm d}}$ data state\footnote{This is a state with entanglement between the data and ancilla qubits, but the data qubits are not entangled with one another.} which evolves to a maximally entangled GHZ state as is expected under the action of the $Z^{\otimes n_{\rm d}}$ Hamiltonian. The gadget can also be observed to process the $X$ observables predictably with period $\pi$. In the subsequent panels, a perturbing single-qubit $X$ term is applied. One sees that the perturbation breaks the symmetry of the Hamiltonian, causing different data qubit observables to evolve heterogeneously at low confinement. However, when the high levels of confinement are used, the data qubit observables again evolve homogeneously, with expected period determined from the scale of the dressed-basis operation in Equation~\ref{sq_driving_overlap}.

This section has thus shown how the chain and subspace terms combine to produce a gapped gadget Hamiltonian with effective interaction $\alpha Z^{\otimes n_{\rm d}}$, and has demonstrated that the intended coherent dynamics emerge once the confinement is sufficiently strong.

\section{Encoded Operators}
\label{sec:encoded_operators}

The goal of this section is to identify physical operators that realize useful dressed-basis operations on the encoded data qubits such that the gadgets can be used in contexts such as adiabatic quantum error correction or whole-minor embedding chain driving.

We wish to apply operations to data qubits in their dressed basis using physical Hamiltonian terms. This requirement is the same as occurs in the use of logical operations applied within quantum error-correcting codes and results from the geometric de-localization of the data-qubit state over ancilla qubits. Therefore, to apply dressed-basis operations, one must find physical operations that have the desired form once transformed by the unitary $\hat U_{\mathrm{enc}}$ and projected into the ground state data-dependent ancillae subspace. In other words, we must not just drive the data qubits between states, but drive the ancilla bits such that they remain in, or with high overlap with, a valid ancilla state as determined by the new data qubit state.

\subsection{Driving data qubits}

In this section, we demonstrate how $X$ operations can be applied to the dressed data-qubits with physical operations. 

Applying single-qubit $X$ operations to the data qubits drives transitions between the dressed states, non-perturbatively, conditional on the overlap between ancilla ground states for the two configurations being non-zero.

The exact quantity we are interested in, then, is the matrix element in which we start in one configuration $\ket{\mathbf{z_1}}\ket{\psi_{\mathbf{z_1},0}}$, and, via the driver, end in state $\ket{\mathbf{z_2}}\ket{\psi_{\mathbf{z_2},0}}$.
Explicitly, this is:
\begin{equation}
    \bra{\mathbf{z_1}}\bra{\psi_{\mathbf{z_1},0}} X_i\ket{\mathbf{z_2}}\ket{\psi_{\mathbf{z_2},0}}.
\end{equation}
Considering the case where the satisfiable allocation $\mathbf{z_1}$ has even parity and an unsatisfiable data state $\mathbf{z_2}$ has odd parity, with them differing by a single bit in position $i$ so $\mathbf{z}_1\oplus 2^i = \mathbf{z}_2$, we can act left with the $X_i$ operator and obtain:
\begin{equation}
\left\langle\psi_{\mathbf{z}_1,0}\middle|\psi_{\mathbf{z}_2,0}\right\rangle.
\end{equation}
The ground state $\ket{\psi_{\mathbf{z}_2,0}}$ consists of a superposition over domain-wall computational basis states for its configuration ${\bf z}_2$ and the ground state $\bra{\psi_{\mathbf{z}_1,0}}$ consists of a single allocation $\mathbf{a}_{\mathbf{z}_1}$ of the ancillae.

\begin{figure*}
    \centering
    \includegraphics[width=\textwidth]{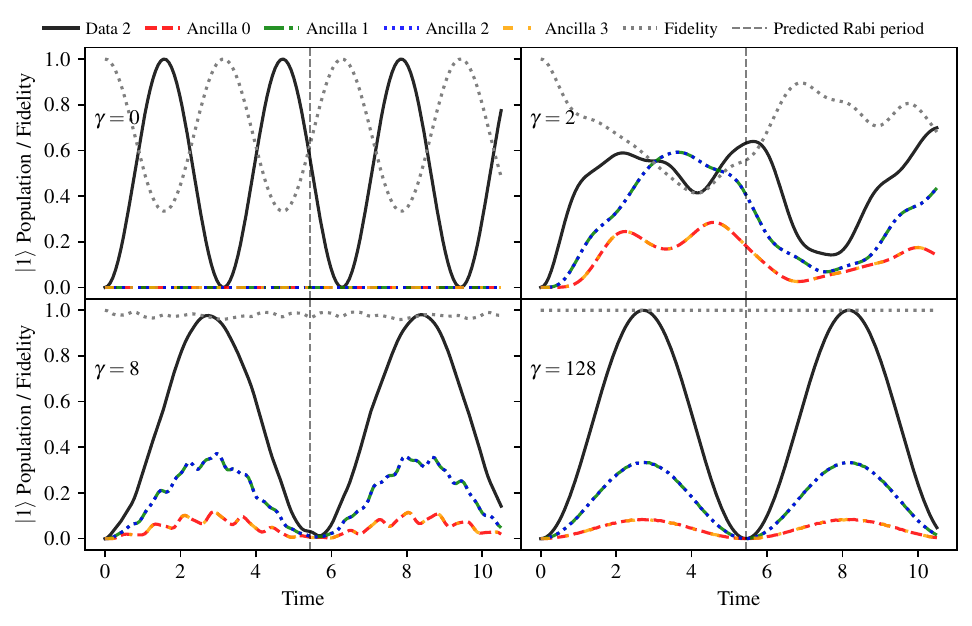}
    \caption{Illustration of bit-flipping dynamics in a $9$-qubit ($n_{\rm{d}} = 5$) gadget system. The figure demonstrates the transitions between dressed states under the influence of a physical single-qubit $X$ driver on the second data qubit and the corresponding response of the ancilla qubits mediated by the gadget constraints. Five data qubits are used, with the system alternating between the dressed states $\ket{00000}$ and $\ket{00100}$, while the ancilla qubits correspondingly alternate between the state $\ket{0000}$ and the domain-wall superposition $0.577\ket{0000} + 0.5\ket{0010} + 0.289\ket{0011} + 0.5\ket{0100} + 0.289\ket{1100}$. Note that due to gauge transforms, the presence or absence of a domain wall cannot simply be determined by agreement or disagreement of neighbouring qubits; the sign of the coupling determined by the data-qubit state must also be taken into account. \label{fig:bit_flipping}}    
\end{figure*}

For the overlap to be non-zero, there should be shared valid ancilla configurations for both data configurations ${\bf z}_1$ and $\mathbf{z}_2$. This is true for bit-flip location $i$ and defect location $j$ only if: 
\begin{equation}
    \prod_{i'=0}^{k} Z_{i'} {(-1)}^{\delta(i,i')}= \prod_{i'=0}^{k} Z_{i'}{(-1)}^{\delta(j,i')}
\end{equation}
for all $k$, which can only be true for $j= i$. So, the overlap is the $j$th component of $\ket{\psi_{\mathbf{z}_2,0}}$, which is:
\begin{equation}\label{sq_driving_overlap}
    {S}_{0 j} = \sqrt{\frac{2}{n_\text{d}+1}}\cdot\sin\left(\frac{\pi (j+1)}{n_{\rm d}+1}\right).
\end{equation}
To act in the dressed basis with a single-qubit $X$ term, we should apply an $X$ operator to the corresponding qubit with a correction $\frac{1}{S_{0j}}$ such that in the dressed basis the terms strength is $1$. The action of a single-qubit driver is demonstrated in Figures~\ref{fig:ghx_and_x_flips} and~\ref{fig:bit_flipping}, in which single-$X$ drivers are applied under different gadget strengths $\gamma $.

If one were to create the gadget in the form of a connected loop, then the diagonalizing operator $\hat S$ in Equation~\ref{sine_transform_eqn} would be a discrete Fourier transform rather than a sine transform. This would result in a ground state with identical elements along the chain and, therefore, uniform sensitivity to driving. If, in the case of an application such as on a minor embedding chain, one wanted increased sensitivity at the ends of the chain rather than at the centre, heterogeneous subspace driving could be used. By applying stronger driving terms towards the ends of the chain than in the centre, one would encourage amplitude to accumulate at the chain ends in the ground state, thereby increasing sensitivity to driving at those locations. This would have the added benefit of reducing sensitivity to spurious driving on the mid-chain, where access would not be needed.

\subsection{Two-body encoded terms\label{sub:two_enc}}

In this section we describe how one can apply two-body terms to the dressed data-qubit states with operations on physical qubits.

In the case of minor embedding~\cite{choi08a} or that of adiabatic quantum error correction~\cite{pudenz2014error,vinci15a}, it is helpful to have access to a logical-$X$ operator for driving within a code, or along the entire minor embedding chain. Alongside these applications, there is also use for such Hamiltonian terms as catalysts for adiabatic quantum computation~\cite{feinstein2024effects, banks2025gadgets}. For use in producing $X^{\otimes n_{\rm d}}$ interactions, the gadget must be used in the Hadamard transformed basis, in which all $Z$ and $X$ observables as previously described are swapped.

In these contexts, it is still needed to penalize breaks on the domain-wall encoded chain brought about by spurious errors or noise. This can be achieved by the usual method of applying ferromagnetic $ZZ$ penalty terms along the chain of qubits one wishes to remain locked to one another. In the working basis of this paper, this corresponds to applying $XX$ terms to data qubits along the chain.

Simply flipping two adjacent data qubits will result in no action on the effective Hamiltonian at zeroth order in perturbation theory, that is:
\begin{equation}
    \hat P\hat U_\mathrm{enc}^\dagger [X_{i}X_{i+1}] \hat U_\mathrm{enc} \hat P = 0.
\end{equation}
A no-defect state, under the action of this operation, experiences the creation of two new defects and therefore there is no overlap in the dressed basis. A one-defect state, on the other hand, will see the defect moved along the chain since the truth values of the two incident clauses are flipped. Since the ancillary state has a different domain wall position, there is no overlap between the dressed states. If, however, we simultaneously flip the ancilla $i$, then the domain-wall position can be simultaneously moved to match the data qubits, and the physical operator has full support on the dressed subspace.

Therefore, we have that:
\begin{equation}
    \hat P\hat U_\mathrm{enc}^\dagger [X_{i}X_{i+1}X_{i}^{\rm a}] \hat U_\mathrm{enc} \hat P = X_{i}X_{i+1},
\end{equation}
and the ability to apply adjacent encoded $XX$ operations in the chain with $3$-body physical terms is apparent.

This section has demonstrated that, while the encoded operators driving transitions between dressed data-qubit states have non-trivial physical forms, one can find physical operators that either act with full support on the dressed subspace or whose action can be corrected by simultaneously driving the ancillae, yielding single-qubit dressed-basis $X$ operations with $1$-body physical terms and adjacent dressed-basis $XX$ operations with $3$-body physical terms.

\section{Performance metrics}\label{section:performance_metrics}

The goal of this section is to define quantitative metrics that separate leakage from in-subspace Hamiltonian error, so that the operating regime of the gadget can be assessed clearly.

\subsection{Leakage}

Leakage quantifies the probability that amplitude leaves the dressed (``valid'') subspace defined by the support of $\hat P$ during time evolution under the gadget Hamiltonian. For the physical propagator \(\hat{O}(t) = {\rm e}^{-it\hat{\mathcal{H}}}\) and the projector \(\hat P\) onto the dressed subspace, we define the survival probability:
\begin{equation}
    p_{\text{surv}}(t) = \frac{\Tr\bigl[\hat P \hat{O}^{\dagger}(t) \hat P \hat{O}(t)\bigr]}{2^{n_d}},
\end{equation}
and hence the leakage:
\begin{equation}
    \text{Leakage}(t) = 1 - p_{\text{surv}}(t).
\end{equation}

\subsection{Conditional Fidelity}

To isolate effective Hamiltonian errors from leakage, we first restrict the physical propagator to the low-energy space and renormalize it for leakage:
\begin{equation}
    \widetilde{O}(t) = \frac{\hat P \hat{O}(t) \hat P}{\sqrt{p_{\text{surv}}(t)}} \quad \bigl(p_{\text{surv}} > 0\bigr).
\end{equation}
Let
\begin{equation}
    \hat{O}_{\text{ideal}}(t) = \exp\bigl[-it \hat P \hat U_{\rm enc}^{\dagger} \hat{\mathcal{H}} \hat U_{\rm enc} \hat P\bigr]
\end{equation}
be the ideal effective propagator acting within the dressed subspace. The conditional process fidelity (post-selected on no leakage) is:
\begin{equation}
    F_{\text{cond}}(t) = \frac{\bigl|\Tr\bigl[\hat{O}_{\text{ideal}}^{\dagger}(t) \widetilde{O}(t)\bigr]\bigr|^{2}}{{(2^{n_{\rm d}})}^{2}}.
\end{equation}
By construction, \(0 \leq F_{\text{cond}} \leq 1\), with \(F_{\text{cond}} = 1\) corresponding to perfect evolution in the dressed subspace. In practice, this means that although amplitude may have left the dressed subspace, it has not returned, carrying phase evolution as a result of time spent in the high-energy sector.

\subsection{Absolute Fidelity}

A single figure of merit that incorporates both failure modes is the absolute fidelity:
\begin{multline}
    F_{\text{abs}}(t) = p_{\text{surv}}(t) F_{\text{cond}}(t) 
    \\
    = \bigl(1 - \text{Leakage}(t)\bigr) F_{\text{cond}}(t),
\end{multline}
which reduces to the survival probability when the dressed-subspace action is exact (\(F_{\text{cond}} = 1\)).

This section has therefore introduced leakage, conditional fidelity, and absolute fidelity as the three complementary diagnostics used in the numerical results and figures that follow.

\section{Performance}
\label{sec:performance}

The goal of this section is to evaluate the gadget in realistic parameter regimes for dressed-basis data driving and for minor-embedding chain flips in the presence of noise.

\subsection{Data driving}

To demonstrate that the ancillae become \textit{locked} to the data qubit under high confinement, Figure~\ref{fig:bit_flipping} shows that the action of driving on the data qubits is to trivially flip their values if no confinement is used. Then, for increasing confinement, the gadget Hamiltonian mediates interactions with the ancillae and causes the ancillae to evolve with the data qubits such that the overall state remains valid.

The period of the oscillations is $\pi$ for the unconfined dynamics, with the non-unit overlap of the dressed states increasing this by a factor of $S_{02}^{-1}$. For five qubits and modest confinement $\gamma = 8$, fidelity with ideal dressed-subspace dynamics remains high.

\begin{figure}
    \centering
    \includegraphics[width=0.49\textwidth,trim=3.5mm 3.5mm 3.5mm 3.5mm,clip]{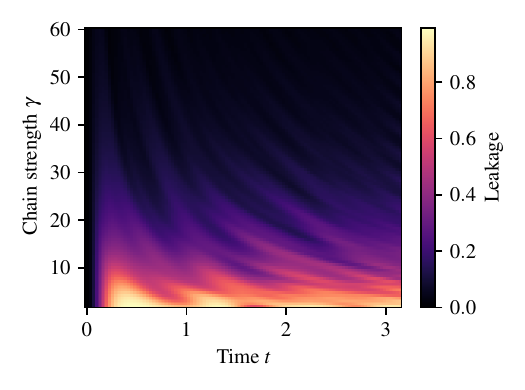}
    \hfill
    \includegraphics[width=0.49\textwidth,trim=3.5mm 3.5mm 3.5mm 3mm,clip]{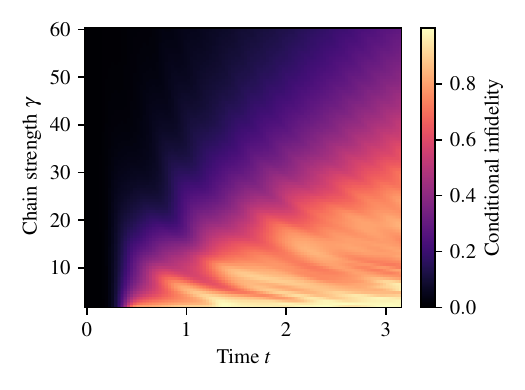}
    \caption{A gadget of five data qubits and four ancilla qubits is tested under the application of a chain of strength $\gamma$, with subspace driving ensuring that a many-body term of unit strength is applied. Data qubits are subject to single-qubit driving with unit strength. (a) Leakage, (b) Conditional infidelity. After some initial mixing, leakage remains at a constant level in time for a given chain strength. Infidelity increases in time for a given confinement strength $\gamma$ due to higher-order leakage processes in which multiple transitions occur in the invalid subspace before a return to the dressed subspace. The total process fidelity is the product of the two fidelities plotted above.\label{fig:gadget_infidelities}}
  \end{figure}

Using the metrics from Section~\ref{section:performance_metrics}, we show in Figure~\ref{fig:gadget_infidelities} the sources of error that accumulate from leakage and the impact of the return of leaked amplitude to the low-energy sector. This shows the expected relationship with the instantaneously leaked fraction remaining constant over time for a given level of containment after some short mixing time, and infidelity accumulating as a function of time for a given level of confinement.

\subsection{Minor embedding}\label{sec:minor_embedding}

\begin{figure}
    \centering
    \includegraphics[width=0.49\textwidth]{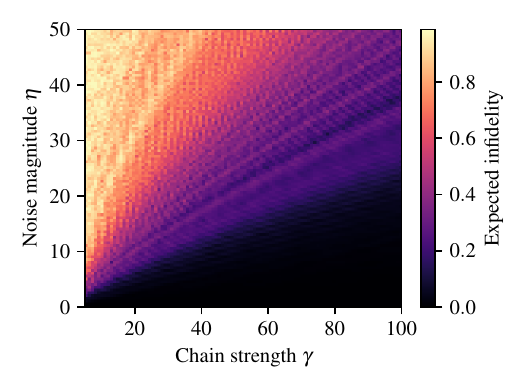}
    \caption{A gadget of five data qubits and four ancilla qubits is initialized in the dressed $\ket{\mathbf{0}}$ state and allowed to evolve under the action of the gadget Hamiltonian in the rotated basis in which it acts as a many-body $X$ operator upon the dressed subspace. Single-qubit $X$ terms are allowed to act on the data qubits to break the minor embedded chain and are suppressed by the addition of logical $ZZ$ terms acting between pairs of data qubits. After time $\pi$, we measure the overlap between the produced state and the dressed $\ket{\mathbf{1}}$ state and plot the associated infidelity for differing chain strength and magnitude of single-qubit terms applied to the data qubits.\label{fig:minor_embedding_infidelity}}
  \end{figure}

Finally, we come to the case of using the gadget in a minor-embedded chain of qubits. Minor embedding uses the gadget in the Hadamard-rotated basis relative to that presented in the rest of the document. Thus, in the opposite basis to the one used above, we start with physical qubits corresponding to the dressed state $\ket{+}^{\otimes n_{\rm d}}$ and act on this with a device Hamiltonian intended to flip the minor-embedding chain. After the application of the device Hamiltonian for a $\pi$ pulse, the device should reach the corresponding dressed state $\ket{-}^{\otimes n_{\rm d}}$. To penalize the chain becoming heterogeneous, and as in the usual case of minor embedding, we apply the physical $3$-body terms $-X_{i}X_{i+1}X_i^{\rm a}$, which implement the logical $-X_{i}X_{i+1}$ couplings and penalize adjacent terms taking differing values. Without some non-commuting term, the resulting dynamics would be trivial, so we add a continuous, time-independent single-qubit-$Z$ error channel on the data qubits with Gaussian-distributed magnitude with scale $\eta$ and zero mean, so $g_i \sim \mathcal{N}(\mu = 0, \sigma = \eta)$. Such a channel can be viewed through the lens of minor embedding as each data qubit being subject to a constant but randomly scaled bit-flipping term. We plot the state fidelity of the end state produced by 
\begin{multline}
\hat O(\gamma,\eta) = \exp(i\pi\hat{\mathcal{H}}_{\rm system}(\gamma, \eta)) 
\\
\hat{\mathcal{H}}_{\rm system}= \hat{\mathcal{H}}_{\rm gadget}(\gamma) + \sum_{i=0}^{n_{\rm d} -1} g_iZ_i\\ + \sum_{i=0}^{n_{\rm d}-2} \gamma X_iX_{i+1}X_i^{\rm a}
\end{multline}
with the predicted ideal end state:    
\begin{equation}
    f \sim {[\bra{-}^{\otimes n_{\rm d}} \hat U_{\mathrm{enc}}^\dagger \hat{O}(\gamma, \eta) \hat U_{\mathrm{enc}} {\ket{+}}^{\otimes n_{\rm d}}]}^2.
\end{equation}
 The expected value of this fidelity is estimated in Figure~\ref{fig:minor_embedding_infidelity}, using an average over $10$ repetitions for each point, and demonstrates parameter regimes in which the noise is either suppressed or fatal for the minor-embedding gadget. Broadly, the plot shows a linear relationship between the magnitude of the applied noise and the strength of confinement required to suppress it.

This section has shown that stronger confinement locks the ancillae to the dressed data dynamics, and that the same confinement scale controls the regime in which the minor-embedding application remains robust against single-qubit noise.

\section{Extensions to More Complex Defect Systems}
\label{sec:extensions}

The domain-wall construction studied in this paper is the simplest example of a system that exploits the odd-even properties of topological-defect subspaces. In the present gadget, domain-wall defects form in an effective Ising spin chain with three-body couplings, with the data qubits setting the signs of the interactions between adjacent sites. More broadly, the dynamics of domain-wall-encoded subspaces is a rich area of study~\cite{werner2025quantum}, allowing particle-like dynamics to be encoded on quantum simulators using local couplings. This makes the present gadget a useful proof-of-principle for a wider class of constructions in which global information is transmitted through constrained defect sectors rather than through explicit high-order physical couplings.

\begin{figure}
    \centering
    \includegraphics[width=0.49\textwidth]{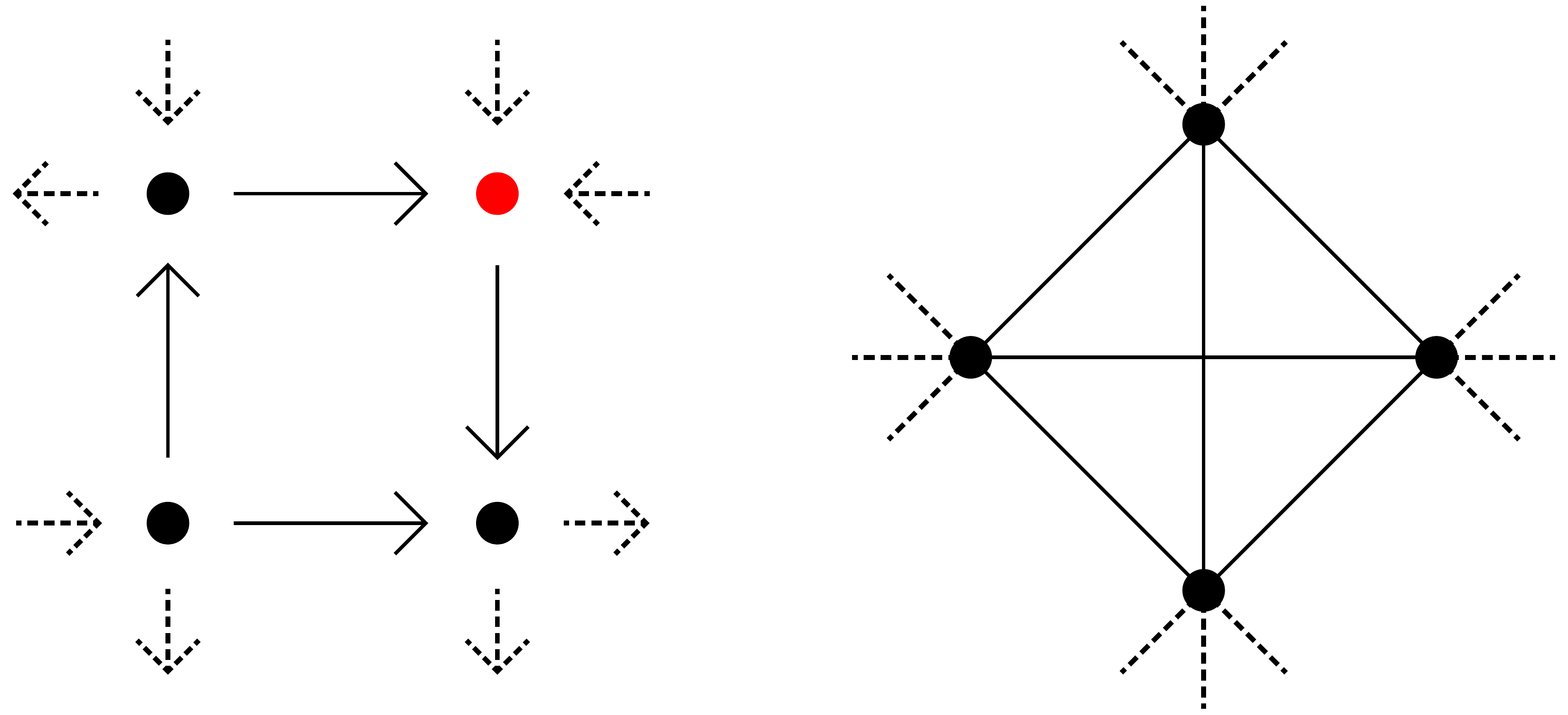}
 \caption{Left: spin-ice systems follow local ice rules. For example, on a tetrahedral pyrochlore spin-ice lattice, a $2$-in $2$-out rule is adhered to, represented here in 2d. Black vertices adhere to the rule and the red vertex does not, as three directed edges flow into the vertex and only one out. Right: the same dynamics can be expressed with anti-ferromagnetic two-body interactions and four spins per vertex of the left-hand diagram.
    \label{fig:spin_ice_fig}}
  \end{figure}

One natural extension is to ice-like systems, where the presence or absence of a topological defect is determined entirely by boundary-spin configurations, even though the bulk interactions are only two-local~\cite{King_2021, wang2006artificial}. Such systems obey local rules that produce extensive global structure through frustration, as illustrated in Figure~\ref{fig:spin_ice_fig}, and their degenerate low-energy sectors support magnetic-monopole-like quasiparticle excitations whose parity depends on the geometry and boundary conditions of the system. Unlike the present domain-wall construction, this suggests a route by which the many-body coupling realized \emph{exactly} here using fifth-order terms might instead be realized \emph{approximately} in a defect-based gadget built from only second-order couplings.

Exploring such systems is beyond the scope of the present work. A realistic study would likely require numerical methods more sophisticated than the matrix diagonalization techniques used here, because even minimal examples would involve substantially larger systems. However, the resulting gadget-plus-chain constructions would remain quasi-one-dimensional, which suggests that tensor-network approaches may provide a viable route for future investigation~\cite{Orus2014tensorNetwork,Berezutskii2025tensorNetworks}.

\section{Conclusion}
\label{sec:conclusion}

Locality-reduction perturbative gadgets are a powerful theoretical tool for showing that high-order terms can be constructed from lower-order ones, but they suffer from serious drawbacks in the number of qubits and energy regimes required for their use. In this work, we have introduced non-perturbative gadgets that produce effective many-body interactions from systems built from three-body chain terms together with either three- or five-body driving terms, with the caveat of complex dressed-basis operations or reduced composability depending on the construction. We demonstrate that the gadget allows for the use of dressed-basis operations needed for a hardware implementation of whole-chain-driven minor embedding. A related use case that would be interesting to explore in the future is self-correcting error-correction codes~\cite{sarovar05a,Jordan06a,young13a,Sarovar13a}. In that case, however, it would be important to design codes that can handle the inherent constraints on composability.

The construction relies on the presence or absence of topological defects determined by the boundary conditions to which they are subject and could form the simplest example of a wider class of constructions for Hamiltonian-based computing in which such properties transmit global information between sites. While this gadget requires three- or five-body terms to implement, other systems such as qubit spin-ice rely only on two-body terms and could yield other interesting constructions as a future area of study. 

\section{Acknowledgements}

This work was funded by the UK EPSRC QCI3 quantum computing hub, grant number EP/Z53318X/1. Figures~\ref{fig:spin_ice_fig} and~\ref{fig:perturb_vs_nonperturb} were illustrated by Robert Headley.

\bibliographystyle{unsrt}
\bibliography{references}

\newpage

\appendix
\section{Perturbative Gadgets}
\label{sec:perturbative_gadgets}
\begin{figure*}
    \centering
    \includegraphics[width=0.9\textwidth]{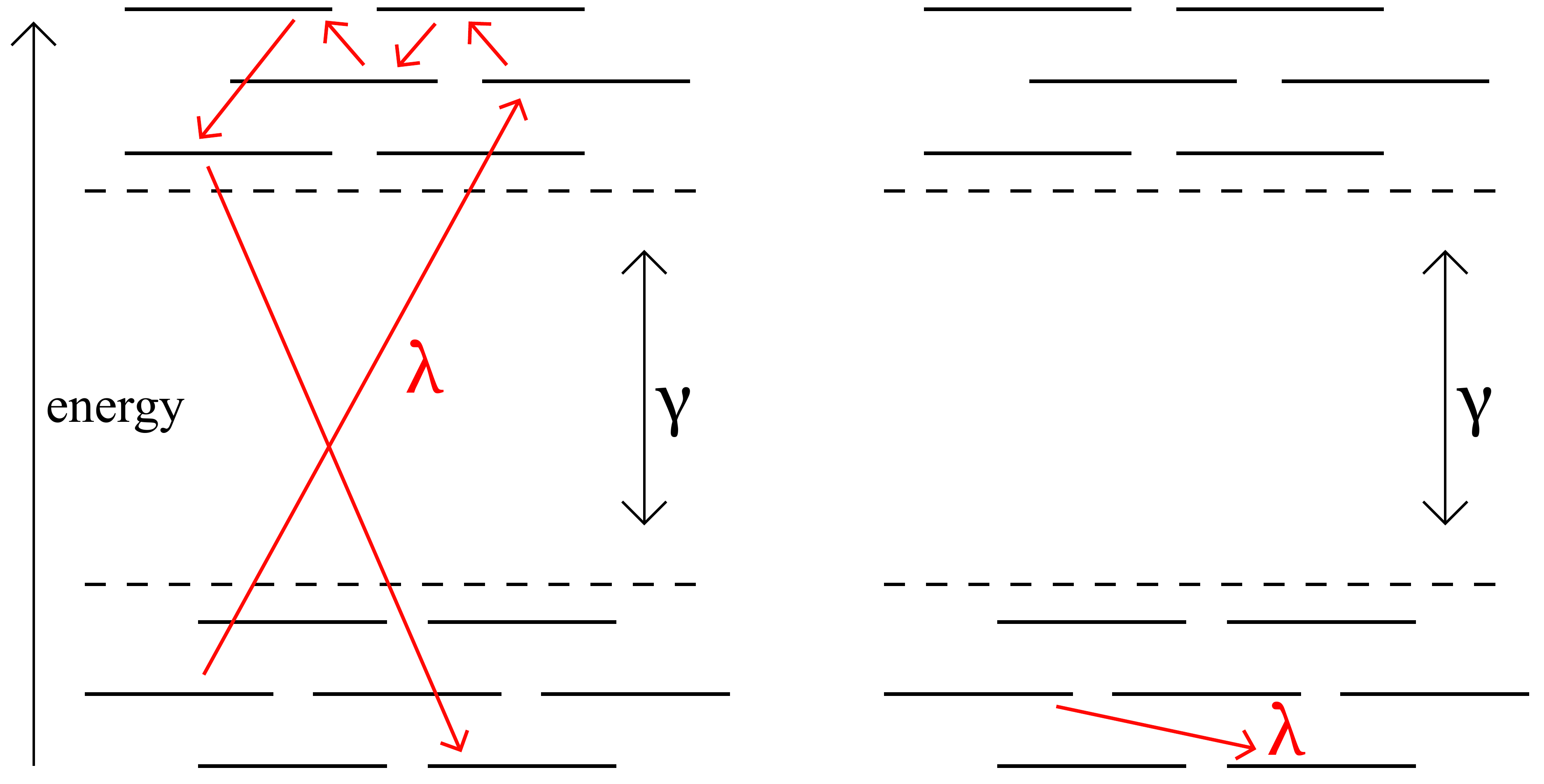}
    \caption{Left: Perturbative gadgets work by penalizing a subspace and applying a perturbation which drives transitions between un-penalized and penalized states. After several \textit{hops} in a penalized space, amplitude may return to the non-penalized space, reaching a configuration or carrying a phase that is not reachable without the path through penalized states. Right: In a non-perturbative gadget the design is such that needed moves between non-penalized states are available without the requirement to take a path through penalized states.\label{fig:perturb_vs_nonperturb}}

\end{figure*}

Quantum gadgets were developed in the context of proving via complexity theoretic arguments that the $2$-local Hamiltonian problem is \textsc{qma}-complete. That is, they form a natural analogue to the classical \textsc{max}-$2$-\textsc{sat} problem on quantum computing devices. It was shown sequentially by Kitaev~\cite{kitaev2002classical} and Kempe~\cite{kempe20033localhamiltonianqmacomplete} that $5$-, and then $3$-local Hamiltonians are \textsc{qma}-complete, but further reductions were needed to demonstrate the $2$-local case. As reductions are used in classical complexity theory to show that problems are \textsc{np}-complete through transformations to problems for which rigorous proofs exist, gadgets that can implement $3$-local Hamiltonians from $2$-local Hamiltonians can be used to show that the dynamics of a $3$-local Hamiltonian can be embedded in that of a larger $2$-local Hamiltonian and, therefore, that the $2$-local Hamiltonian problem is also \textsc{qma}-complete~\cite{kempe2006complexity}. These perturbative gadgets were designed primarily as theoretical, not practical, tools. The difference in energy scales between the produced higher-order interactions and the physical, low-order interactions used to generate these, alongside the per-interaction requirement for ancillary qubits, limits the practicality of the constructions. By contrast, the non-perturbative gadgets presented in this work rely on direct transitions between valid dressed states rather than paths through penalized subspaces, as illustrated in Figure~\ref{fig:perturb_vs_nonperturb}.

Perturbative gadgets are typically constructed through the use of a large energy scale penalizing states as to define a low-energy subspace. However, this penalty must generally not be absolute, since the functioning of gadgets depends on the ability to pass through penalized states to transit between un-penalized states. Using the dressed-basis formalism introduced in the main text, the effective Hamiltonian within the low-energy subspace is obtained order by order via the Schrieffer-Wolff transformation~\cite{schrieffer1966relation}. A gadget is deemed \textit{perturbative} at $k$th order when one must consider $k$ transitions, or powers of a perturbing Hamiltonian which drives transitions between  penalized and un-penalized subspaces, to determine the relevant effective dynamics. For example, in a paper demonstrating gadgets that reduce terms of arbitrary order to two-local terms~\cite{jordan2008perturbative}, $k$ ancillae with all-to-all ferromagnetic coupling of strength $\gamma$ are added to $k$ data qubits. Driving the data-ancillae system with terms of the form
\begin{equation}
    \hat{\mathcal{H}}_\mathrm{drive} = \lambda\sum_{i=0}^{n_{\rm d}-1} Z_i X^{\rm a}_i
\end{equation}
gives dynamics in which each term in the drive takes the system outside a valid space defined by the even GHZ ancilla cat state. It is only after raising $\hat{\mathcal{H}}_\mathrm{drive}$ to the $k$th power that one obtains a non-trivial path from the valid ancilla state ($\prod_i X_{\rm a}^i$ acts trivially on the cat state, but one accumulates each factor of $Z_i$ on the data qubits), back to itself, and, therefore, one must apply perturbation theory at $k$th order to obtain a non-trivial effective Hamiltonian. The effective interaction produced by this gadget scales as $O\left(\frac{\lambda^k}{\gamma^{k-1}}\right)$, but for the perturbative expansion to be valid, one needs $\gamma \gg\lambda$. The requirements of needing strong interactions and also strong confinement are, therefore, in conflict, resulting in the need for energy scales for both penalty and perturbation both scaling exponentially in the order $k$ of perturbation theory used.

\end{document}